\documentstyle[aps]{revtex}

\newcommand{\bbf}[1]{\mbox{\boldmath{$#1$}}}
\newcommand{\bean}{\begin{eqnarray}}
\newcommand{\eean}{\end{eqnarray}}
\newcommand{\bea}{\begin{eqnarray*}}
\newcommand{\eea}{\end{eqnarray*}}
\newcommand{\ben}{\begin{equation}}
\newcommand{\een}{\end{equation}}
\newcommand{\be}{\begin{displaymath}}
\newcommand{\ee}{\end{displaymath}}
\newcommand{\ba}{\begin{array}}
\newcommand{\ea}{\end{array}}
\newcommand{\kn}{\nonumber}

\newcounter{saveeqn}

\newcommand{\alpheqns}{\setcounter{saveeqn}{\value{equation}}%
\setcounter{equation}{0}%
\renewcommand{\theequation}{\arabic{saveeqn}\alph{equation}}}
\newcommand{\reseteqn}{\setcounter{equation}{\value{saveeqn}}%
\renewcommand{\theequation}{\arabic{equation}}}

\begin{document}

\title{Mesoscopic motion of atomic ions in magnetic fields}
\author{David M.~Leitner}
\address{Department of Chemistry, University of Illinois at Urbana-Champaign,\\
Urbana, Illinois 61801}
\author{P. Schmelcher}
\address{Theoretische Chemie, Physikalisch-Chemisches Institut, Universit\"at 
Heidelberg, Im Neuenheimer Feld 253, 69120 Heidelberg, Germany}

\date{\today}
\maketitle

\begin{abstract}
We introduce a semiclassical model for moving highly excited atomic ions
in a magnetic field which allows us to describe the mixing of the
Landau orbitals of the center of mass in terms of the electronic excitation
and magnetic field. The extent of quantum energy flow in the ion is investigated
and a crossover from localization to delocalization with increasing center of mass energy
is detected. It turns out that our model of the moving ion in a magnetic field is closely 
connected to models for transport in disordered finite-size wires.
\end{abstract}




Interacting particle systems in strong magnetic fields show a rich variety
of complex phenomena. The source of this complexity is competition between the
magnetic and Coulomb interactions which are of inherently different character.
With changing strength of the external field the
corresponding systems undergo a metamorphosis involving qualitatively different
states. In atomic physics attention focused for more than a decade on the
hydrogen atom in a magnetic field (see ref.\cite{Fri89} and references therein)
on which detailed experimental and theoretical investigations
yielded many beautiful insights into semiclassical and quantum aspects of 
nonintegrable systems and significantly enhanced our understanding of the new 
features arising due to the presence of the external field. 
With increasing degree of excitation and/or increasing field strength the
electronic motion of the classical atom shows a transition from regular to irregular, i.e.
chaotic, behavior and intermittency.
More recently it has become evident that the non-separability
of the collective, i.e., center of mass (CM), and electronic motion of atoms in the presence
of a magnetic field leads to a variety of two-body phenomena. 
The corresponding coupling of the CM and electronic motion is fundamentally different for
neutral and charged systems.
Examples of two-body effects in neutral systems are the classical chaotic diffusion
of the CM \cite{Schm92} or the existence of weakly bound giant dipole states \cite{Bay92}. 
For atomic ions the interaction of the CM and electronic degrees of freedom is
more intricate and manifests itself in a continuous, classical flow of energy from the
collective to the internal motion and vice versa. Detailed studies of the classical dynamics
of rapidly moving highly excited $He^{+}$-ions in a magnetic field showed 
that this energy exchange leads to the self-ionization process \cite{Schm95} of the ion.
Very little, however, is known about the quantum properties and behavior
of moving highly excited atomic ions. 

Using Landau orbitals for the CM motion in zeroth-order
and fixed nucleus zero-field wave functions for the electronic motion
to estimate their coupling matrix elements, it was demonstrated \cite{Schm91} that there exist
a number of different physical situations for which the interaction between
the collective and electronic motion becomes strong. The latter induces a strong
mixing of the CM and electronic motion and is a potential source of
interesting quantum properties of the ion, in particular when the dynamics of the corresponding
classical ion is chaotic.  A detailed investigation of
the coupled CM and electronic motion of the highly-excited quantum mechanical
ion in this regime, which
is the subject of interest of the present paper, is
however a highly nontrivial task: we are dealing with five nonseparable
and strongly mixing degrees of freedom in a regime of very high level density which depends
on a number of parameters (field strength, total energy, etc.). 
The ab initio description of the quantum dynamics in the above regime goes even beyond
modern computational possibilities and we thus seek a model approach
that captures the essential physics of the problem.
Here we propose and analyze a semiclassical model of the excited ion, and explore
the consequences of coupling between its CM and electronic degrees of freedom.

Since we deal with the interaction of the CM and electronic motion in atomic ions
we first have to introduce collective (CM) and relative variables in the Hamiltonian
describing the atom. The total pseudomomentum ${\bbf{K}}$ \cite{Avr78} is a conserved quantity
associated with the CM motion which, in spite of the fact that its components
perpendicular to the magnetic field are not independent, i.e. do not commute,
can be used to transform the Hamiltonian to a particularly simple and physically appealing
form \cite{Schm91} which for the $He^{+}$-ion reads
${\cal H} = {\cal H}_1 + {\cal H}_2 + {\cal H}_3$ where
\setcounter{equation}{1}
\alpheqns
\ben\label{Hamiltonian}
{\cal H}_1 =  \frac{1}{2M}\left({\bbf{P}}-\frac{Q}{2}{\bbf{B}}\times
{\bbf{R}}\right)^2 
\een
\ben
{\cal H}_2 = \alpha \frac{e}{M}\left({\bbf{B}}\times\left({\bbf{P}}-\frac{Q}{2}{\bbf{B}}\times
{\bbf{R}}\right)\right){\bbf{r}}
\een
\ben
{\cal H}_3 = \frac{1}{2m}\left({\bbf{p}}-\frac{e}{2}{\bbf{B}}\times
{\bbf{r}}+\frac{Q}{2}\frac{m^2}{M^2}{\bbf{B}}\times{\bbf{r}}\right)^2 \\
+\frac{1}{2M_0} \left({\bbf{p}}+\left(\frac{e}{2}-\frac{Q}{2M}\frac
{m}{M}\left(M+M_0\right)\right){\bbf{B}}\times
{\bbf{r}}\right)^2 -\frac{2e^2}{r}\kn
\een
\reseteqn
where $m,M_0$ and M are the electron, nuclear and total mass, respectively.
$\alpha=(M_0+2m)/M$ and $Q$ is the net charge of the ion. ${\bbf{B}}$ is the
magnetic field vector which is assumed to point along the z-axis.
$({\bbf{R}},{\bbf{P}})$ and $({\bbf{r}},{\bbf{p}})$ are the canonical pairs
for the CM and internal motion, respectively. ${\cal H}$
involves five degrees of freedom since 
parallel to the magnetic field the CM
undergoes free translational motion and can be separated completely.

${\cal{H}}_1$ and ${\cal{H}}_3$ depend exclusively on the
CM and electronic degrees of freedom, respectively. ${\cal{H}}_1$ describes the
free motion of a CM pseudoparticle with charge Q and mass M.
${\cal{H}}_3$ describes the electronic motion in the presence of paramagnetic,
diamagnetic as well as Coulomb interactions which, in analogy to the hydrogen
atom \cite{Fri89}, exhibits an enormous complexity of classical and quantum
properties with changing parameters, i.e. energy and/or field strength.
${\cal{H}}_2$ contains the coupling between the CM and electronic motion of the
ion and represents a Stark term with a rapidly oscillating electric field
$\frac{1}{M}\left({\bbf{B}}\times\left({\bbf{P}}-\frac{Q}{2}{\bbf{B}}\times
{\bbf{R}}\right)\right)$ determined by the dynamics of the ion. It is the
interaction Hamiltonian ${\cal{H}}_2$ which is responsible for the interesting quantum
effects that will be investigated in the present Letter.

>From the above it is natural to consider the representation of the coupling
Hamiltonian ${\cal{H}}_2$ in a basis which consists of products of eigenstates 
$\Phi_{CM}$ of ${\cal{H}}_1$ and $\Psi$ of ${\cal{H}}_3$. Calculating the corresponding
matrix elements we encounter some selection rules which are of immediate
relevance to our model (see below). Since the total angular momentum component parallel 
to the magnetic field ${\cal{L}}_z$ is a conserved quantity for ${\cal{H}}$ and since 
the corresponding CM angular momentum $L_{CM_{z}}$ and electronic angular momentum $L_z$ 
are conserved quantities for ${\cal{H}}_1$ and ${\cal{H}}_3$, respectively, the
matrix elements of ${\cal{H}}_2$ involve only CM and electronic
states with magnetic quantum numbers which are correspondingly different.
In addition we have the relation $<{\Phi}_{CM}'|\left({\bbf{P}}-
\frac{Q}{2}{\bbf{B}}\times {\bbf{R}}\right)|\Phi_{CM}> = iM(E'-E)<{\Phi}_{CM}'|{\bbf{R}}|
{\Phi}_{CM}>$ which, together with the dipole selection rules for electronic transitions,
allows only changes of the CM ($\mu$) and electronic ($m$) magnetic quantum numbers by one and
requires a change of energy for the CM motion. According to ref.\cite{Schm91}
the matrix elements of ${\cal{H}}_2$ involve a factor $\sqrt{N}$ for $N>>|\mu|$, ($N$ is
the Landau principal quantum number of the CM motion)
which yields the scaling of the coupling matrix elements of ${\cal{H}}_2$
with respect to varying CM energy.

Our model for the moving $He^{+}$-ion in a magnetic field is built up from
three key constituents associated with the Hamiltonians ${\cal{H}}_1,{\cal{H}}_2$ and
${\cal{H}}_3$. The equidistant and infinitely degenerate spectrum of ${\cal{H}}_1$ is
completely characterized by the CM quantum numbers $N$ and $\mu$. ${\cal{H}}_1$ represents 
the integrable part of the system which is
coupled via ${\cal{H}}_2$ to the chaotic part represented by ${\cal{H}}_3$.
The classical dynamics of ${\cal{H}}_3$ \cite{Del84} depends on the scaled energy and angular momentum
which are given by $\hat{L}_z=L_z \left(\frac{1}{4} B\right)^{\frac{1}{3}}$
and $\hat{E}=E\left(2B\right)^{-\frac{2}{3}}$, respectively.
In order to locate the regime of chaotic electronic motion we have made Poincar{\' e}
surfaces of section (PSOS) of the classical dynamics of ${\cal{H}}_3\left(\hat{E},\hat{L}_z\right)$
for a dense grid of values of the scaled energy and angular momentum.
Starting with a completely chaotic phase space for $\hat{L}_z=0$ 
we find that with increasing values of the angular momentum the fraction of chaotic phase
space volume decreases rapidly. In contrast, the regime of negative $\hat{L}_z$
that yields predominantly chaotic phase space is much larger. If we take a typical scaled energy of $\hat{E}=-0.1$,
for example, and require more than $90\%$ of phase space to be occupied by chaotic trajectories
we obtain the regime $\hat{L}_z=[-2.71,0.136]$ which corresponds to $L_z=[-200,+10]$
for $B=10^{-5}$ (we use atomic units throughout, i.e. the field strength $B=1$ a.u. corresponds
to $2.35 \cdot 10^{5}$ Tesla). The mixing of electronic eigenfunctions belonging to negative values
of the angular momentum $L_z$ represents therefore an important 'open channel' with respect to the coupling
of the chaotic electronic motion to the CM motion.
In this region of $L_z$ the CM motion of the classical ion is strongly affected by
coupling to the chaotic electron.
Indeed, studies of the classical dynamics of the ion close to the
ionization threshold demonstrated \cite{Schm95} that large negative values of the angular momentum
are an inherent feature of intermittent dynamics as well as a prerequisite for
the self-ionization process, whereby energy transfer from the CM to the electron
results in ionization.
We therefore turn our attention to this subspace and investigate the quantum
mechanical properties of this channel.

The spectrum belonging to the chaotic Hamiltonian ${\cal{H}}_3$ will be 
represented by a random matrix ensemble, which is the appropriate semiclassical
description \cite{btu93,lkc94}.  Since the Hamiltonian ${\cal{H}}_3$
possesses a generalized time reversal invariance \cite{Haak91}, which consists of
a rotation by $\pi$ around the $x$-axis and a subsequent conventional time reversal
operation, the proper ensemble is the Gaussian Orthogonal Ensemble (GOE).   
While the GOE provides the
fluctuations of the chaotic levels, we still need to specify the
mean level density (MLD) as a function
of the energy, field strength and in particular the angular momentum $L_z$.
Our approach to the MLD is via the semiclassical
Thomas-Fermi formula. Performing the appropriate scale transformations, fixing $\hat{L}_z$,
transforming to cylindrical coordinates and subsequently performing the integrations over
$\phi,p_z,p_{\rho},z$ we arrive at the following result for the semiclassical MLD
\begin{equation}\label{density}
\rho_{L_z}(E,B)= \left(2B^{-2}\right)^{\frac{2}{3}} 2 \int\limits_{0}^{\infty} d\rho~
\sqrt{\cal{A}}~\Theta(\cal{A})
\end{equation}
with ${\cal{A}} =\left(\frac{1}{2\rho^2}\left(\frac{\rho^2}{2}+\hat{L}_z\right)^2-\hat{E}\right)^{-2} -\rho^2$.
The remaining integration over $\rho$ has to be performed numerically. 
Starting with $L_z=0$ a general feature
of the MLD is its rapid decrease with increasing $L_z$ whereas for negative values of $L_z$ its decay
is much weaker. Not only the fraction of chaotic phase space but also the absolute phase space
volume persists down to large negative values of the angular momentum, to 
$L_z\approx -250$, and therefore $\rho_{L_z}(E,B)$
represents in this regime the density of irregular states. Having specified our GOE,
whose MLD at the center of the band is given by eq.(2) 
providing the levels of ${\cal{H}}_3$, we turn to the calculation of the coupling matrix elements
introduced by ${\cal{H}}_2$. The size of the matrix elements of ${\cal{H}}_2$ can be determined
from a semiclassical relation between off-diagonal matrix elements of an operator, and the
Fourier transform of its classical autocorrelation function \cite{Fein86}. The variance $\sigma_{{\cal{H}}_2}^2$
of the matrix elements of ${\cal{H}}_2$ depends on the energies of the states they couple and is very
small when the energy difference is greater than the level spacing of ${\cal{H}}_1$ \cite{Fein86}.
For states close in energy, $\sigma_{{\cal{H}}_2}^2$ appears as
\begin{equation}\label{offdia}
\sigma_{{\cal{H}}_2}^2 \approx \left(4 \pi \rho_{L_z} \right)^{-1} \int_{-\infty}^{\infty}
<{\cal{H}}_2(t){\cal{H}}_2(0)> e^{-i\omega t} dt,  
\end{equation}
where selection rules determine the numerical coefficient of eq.(3)  \cite{lkc94}.
Efficient evaluation of the above equation through classical trajectory calculations is by no means
trivial and is done by introducing a suitable ensemble average via the periodogram approximation \cite{Daven58}.
Fig. 1 shows  $\chi\equiv\rho_{L_z} |\sigma_{{\cal{H}}_2}|$, as a 
function of the angular momentum, calculated using eqs.(1-3).
As examples we plot $N$=250, 300 and 400 
(at $m=0$) and $\mu<0$.  Due to the selection rules $N$
increases (decreases) by 1 as $m$ decreases (increases) by 1.
As seen in the figure, the largest values of 
$\chi$ lie mainly in the interval $m\approx[-20,-200]$.

Based on the structure of ${\cal{H}}$
the organization of our model is as follows: 
We have an array of sites
each of which corresponds to eigenstates of ${\cal{H}}_1$ and
${\cal{H}}_3$ labeled by particular values of $N$, $m$ and $\mu$. 
We thus assign to each site the CM quantum numbers $N,\mu$ of
a Landau orbital belonging to ${\cal{H}}_1$, and the levels of one member of the GOE representing
the irregular levels of ${\cal{H}}_3$, which are labeled by $m$.  
The energy levels at each site are
the ${\cal{H}}_3$ levels plus the level of ${\cal{H}}_1$.
They are coupled by random matrix elements to levels of their neighboring sites,
as imposed by the selection rules.  The variance of the random elements
$\sigma^2_{{\cal{H}}_2}$ is given by eq.(3).
The model thus has a one-dimensional structure, where sites labeled by
$N$, $m$ and $\mu$ are comprised of levels taken to be from a GOE 
whose density is given by eq.(2).  The ion model so defined resembles 
the semiclassical pump model
of Arnold diffusion \cite{Leit97a}.  The classical stochastic pump model 
describes Arnold diffusion in terms of pumping of otherwise
regular trajectories via weak, irregular motion within the Hamiltonian system \cite{Lich92}.  
Similarly, the classical CM motion of the ion is coupled to the chaotic motion
of the electron due to the magnetic field.  The semiclassical 
pump model of Arnold diffusion was found to be equivalent to
models of single-particle transport in disordered wires, which
predict localization \cite{Thou77,Efet83,Pich86}.	 
The localization length of the semiclassical pump model
was expressed in terms of corresponding classical parameters.
Likewise, we should be able to predict any localization of the quantum ion in terms
of the corresponding classical system.

The values of all the ion parameters
$(N,\mu,E_3,L_z,B)$ are embodied in the model parameter $\chi\equiv\rho_{L_z} 
|\sigma_{{\cal{H}}_2}|$, examples of which are shown in Fig. 1.  Though $\chi$ clearly
fluctuates over the sites of the model, we could nevertheless estimate the localization
length for this model in terms of an average, $\bar\chi$, over $\chi$.  As seen in Fig. 1, the range in $m$
over which energy transfer can most appreciably occur is $m\approx[-20,-200]$,
though the model also encompasses a wider range of $m$, where classically the
ion is still predominately chaotic.  In terms of the average $\bar\chi$,
the localization length, $\xi_\infty$, 
of the model assuming an
infinite number of sites, is $\xi_\infty = 4\pi^2{\bar\chi}^2$ \cite{Leit97a,Efet83}.
Because the ion model has a finite length of about 200 sites, the localization
length can be estimated using finite-size scaling arguments
for band random matrices, like the ion model,
as $\xi \approx \xi_\infty/(1 + c\xi_\infty/L)$, $c\approx 1$  \cite{cim90}, 
where $L$ is the length, or number of sites of the model.  The $N$ dependence of
$\xi$ arises from $\xi_\infty\sim N$, since $|\sigma_{{\cal{H}}_2}|\sim \sqrt{N}$
(see above).  Solving for $\rho_{L_z}$ and
$|\sigma_{{\cal{H}}_2}|$ using eqs.(2) and (3) we find, e.g.,
that $\xi\approx\xi_\infty\approx 30$ for $N$=400 ($m$=0).  Thus 
starting in a Landau level near $N\approx 400$, quantum flow is restricted
to about 30 sites in $m$.  This is in contrast to the classical ion, where
for a corresponding initial CM energy of $\approx 6\times 10^{-7}$ a.u.
there is no such restriction in
the chaotic motion over $L_z$.  Since $\chi$ varies as $\sqrt{N}$ the
ion remains localized, in contrast to the classical ion, for $N$ up to values
near 4000.  Taking the cyclotron frequency to be $1.4\times 10^{-9}$ a.u., this 
corresponds to a CM energy of about $6\times 10^{-6}$ a.u.

These arguments, while using well known predictions for equivalent 
random matrix models, nevertheless depend on our being able to use an average of $\chi$ over the
length of the model to estimate $\xi$.  As a check, we have
studied numerically the ion model to compare predictions
of $\xi$ using the actual $\chi$ which varies as determined
by the semiclassical results for the ion calculated with eqs.(1-3),
with results using $\bar\chi$, defined as the average $\chi$ over the
length of the model.
Our numerical model ranges from $m$=-230 to
0.  Each site consists of 12 levels of the GOE.  Though this is a very
small number, it is all we could include computationally while incorporating also
the largest possible number of sites.  We average over 6 realizations of
each set of parameters and calculate the localization length of the 
eigenvectors.  Results are shown in Fig. 2, where  
we see that both models, the ion and the simpler version with $\bar\chi$, 
give the same results for $\xi$, and are close to 
the line $\xi=0.72\xi_\infty$, where $\xi_\infty=4\pi^2{\bar\chi}^2$, and the factor 0.72 is purely an
artifact of using only 12 levels per site, which is seen upon comparing with numerical
results where more levels per site were used \cite{Leit97a}.  

The occurence of the crossover from localized to delocalized CM motion that we find for the ion at a CM energy
of about $6\times 10^{-6}$ a.u. with the above parameters will of course vary with the
strength of the field and the internal energy of the ion, as well as its mass, since
we have here considered only $He^+$.  
This crossover should be observable spectroscopically since the regime
of very weak and strong mixing show inherently different level spacings and
absorbtion features.  Finally, we mention that
the above-investigated quantum mixing of collective and electronic motion
for atomic ions is certainly of interest also for charged molecular systems
in a magnetic field. Here the heavy vibrational and rotational degrees of
freedom couple, for heteronuclear systems, to the collective motion of the
molecular ion providing a potential source of new rotational and
vibrational structures.

This work was completed during a visit of P.S. at the Max-Planck Institute
for Physics of Complex Systems which is gratefully acknowledged for its 
hospitality.  DML acknowledges support from NSF CHE 95-30680.

\vskip 1.0cm
\noindent
Figure 1.  The model parameter $\chi\equiv\rho_{L_z} |\sigma_{{\cal{H}}_2}|$, 
where $\rho_{L_z}$ and
$|\sigma_{{\cal{H}}_2}|$ are calculated using eqs.(2) and (3), respectively, as a function
of the angular momentum quantum number, $m$.  At $m$ = 0, $N$ = 250
(short dashes), 300 (long dashes) and 400.
\vskip 0.5cm
\noindent
Figure 2.  Localization length computed for the ion at different values of
the Landau level, $N$, at $m$=0, from which $\chi$ is computed with
eqs.(2) and (3).  Filled circles are results for the
ion model and open circles for simpler version described in text. 
The line is a fit through the data.



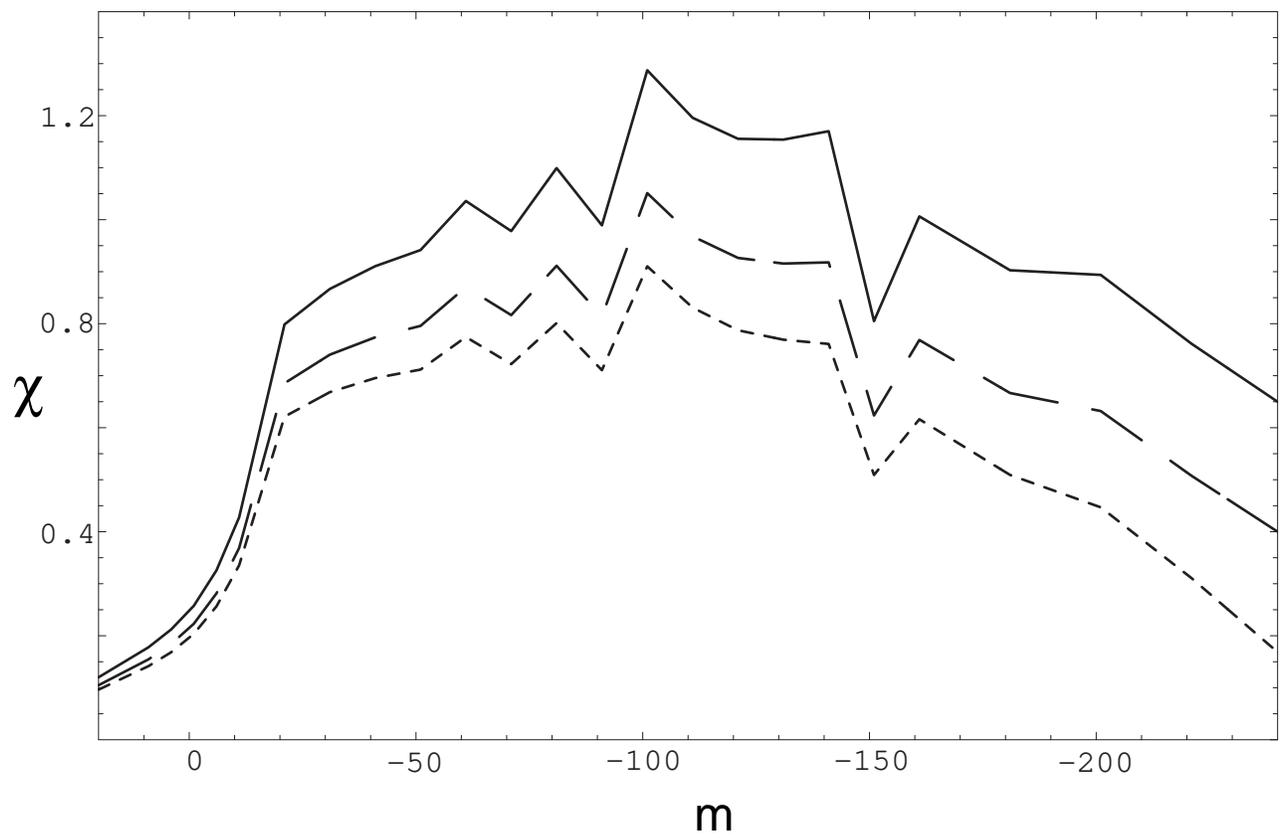
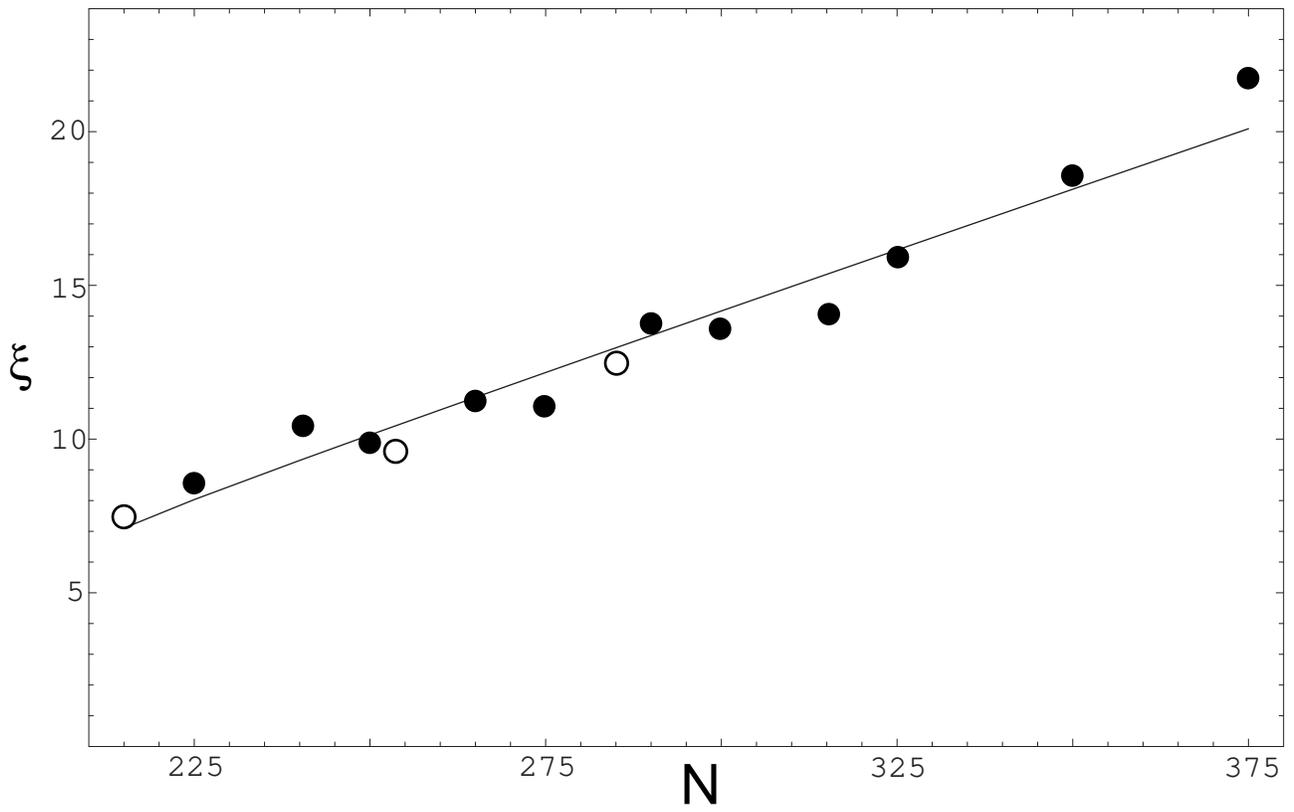
\end{document}